# Single-electron tunneling PbS/InP neuromorphic computing building blocks

P. F. Jarschel[1,2,*], J. H. Kim[1], L. Biadala[3], M. Berthe[3], Y. Lambert[3], R. M. Osgood[4], G. Patriarche[5], B. Grandidier[3], Jimmy Xu[1]

1. *School of Engineering, Brown University, Providence, Rhode Island 02912, USA*
2. *"Gleb Wataghin" Physics Institute, Univ. of Campinas, 13083-859 Campinas, SP, Brazil*
3. *Univ. Lille, CNRS, Centrale Lille, ISEN, Univ. Valenciennes, UMR 8520 - IEMN, F-59000 Lille, France*
4. *US Army Combat Capabilities Development Command – Soldier Center, 15 General Greene Ave., Natick, MA 01760, USA*
5. *Centre de Nanosciences et de Nanotechnologies (C2N), UMR 9001 CNRS, University Paris Sud, University Paris-Saclay, avenue de la Vauve, 91120 Palaiseau.*

## Abstract

We study single-electron tunneling (SET) characteristics in crystalline PbS/InP junctions, that exhibit single-electron Coulomb-blockade staircases along with memory and memory-fading behaviors. This gives rise to both short-term and long-term plasticities as well as a convenient non-linear response, making this structure attractive for neuromorphic computing applications. For further insights into this prospect, we predict typical behaviors relevant to the field, obtained by an extrapolation of experimental data in the SET framework. The estimated minimum energy required for a synaptic operation is in the order of 1 fJ, while the maximum frequency of operation can reach the MHz range.

## 1. Introduction

Single-electron tunneling (SET) has been studied for some time, partly due to its potential for ultra-low energy applications, such as single-electron transistors [1] [2] [3]. Having been observed in many different structures and materials, the SET effect is being explored for many other applications where single charge resolution is required [4] [5], and is particularly relevant to the study of junctions in nanoparticles or quantum dots and boxes (Coulomb islands), where the Coulomb Blockade effect limits the tunneling of single electrons [6] [7] [8] [9].

Due to its intrinsically non-linear response, low energy demand and fast operation, one application of particular interest to SET devices is neuromorphic computing [10] [11] [12] [13]. In it, instead of the traditional von Neuman linear architecture, a bio-inspired (neural) network of non-linear devices and weighting elements is responsible for data processing, allowing computation orders of magnitude faster, as demonstrated by IBM's TrueNorth [14], Stanford's Neurogrid [15], Heidelberg University's HICANN [16], and University of Manchester's SpiNNaker project [17]. This opens the possibility of overcoming the present limitations faced by Moore's law and Dennard scaling [18]. One characteristic of great importance for neuromorphic computing applications is the plasticity, related to memory effects in both short and long-term regimes, for which many resort to memristive devices [19]. If the conductive state of a certain device is affected by previous signals, changing its response to subsequent stimuli, self-learning is possible, increasing the similarities to biological processing, providing even more enhanced data processing capabilities [20].

SET has been previously demonstrated in many materials and structures, one of which is lead sulfide (PbS) [21], a material that recently has been shown to enhance the switching parameters of memristive devices [22], finding its importance to the neuromorphic computing field. In this work, we propose the application of liquid-phase grown PbS nanoplatelets on InP for single-electron neuromorphic computing applications. We demonstrate the occurrence of SET in the self-formed oxide interface between the two materials. From the experimentally observed short and long-term memory effects, with respective time scales in the order of seconds and hundreds of seconds, we extrapolate a SET model to show short and long term plasticity, demonstrating synaptic depression effects. The estimate of the minimum energy required for synaptic operations in such proposed devices is of the order of 1 fJ, equivalent to a minimum pulse width of 200 ns, or a maximum operating frequency of 5 MHz. Comparatively, the human brain synapses consume more energy (~pJ) and operate at a slower rate (1 ms or 1 kHz) [23], which demonstrates the potential of the proposed structure for such applications.

## 2. Fabrication and characterization

The samples are made of PbS nanoplatelets (NPs) grown on top of InP. To obtain these NPs, first a highly polished p-type InP(001) (AXT Inc) substrate is treated in a 1% HF solution, to remove the native oxide. Subsequently, this substrate is immersed in a solution of 1 mM of lead acetate (Sigma), 1 mM of thiourea (Sigma), 10 ml of ethanol (Sigma), and 10 ml of ethylene glycol (Sigma) at 95 ˚C for 1 hour. The pH of the solution was adjusted to 3.4 by adding glacial acetic acid. Finally, the sample was washed with ethanol and dried with a nitrogen blower.

This procedure resulted in the growth of PbS nanoplatelets on the InP substrate as shown in Figure 1(a), predominantly consisting of a square top facet, consistent with the cubic rock-salt structure of PbS. Their lateral sizes vary between 40 and 140 nm, substantially smaller than the biological neurons. The existence of a linear correlation between the height of the NPs and their lateral sizes can be verified by atomic force microscopy [24].

The interface between PbS and InP was studied by combining cross-sectional high angle annular dark field scanning transmission electron microscopy (HAADF-STEM) with energy dispersive spectroscopy (EDS). To this end, thin lamellae were cut from the samples by focused ion beam and deposited on carbon-coated copper grids and analyzed in a JEOL 2200FS microscope operating at 200 KeV. Figure 1(b) shows a clear dark interfacial layer between the substrate and the NP, for a sample prepared parallel to the {100} edge facets. The observation of the interface at higher magnification (Figure 1(c)) shows the lack of a crystallographic structure in the dark layer, indicating the existence of an amorphous layer between the InP substrate and the PbS NP. The presence of the interface is further evidenced in Figures 1(d), 1(e), and 1(f), where EDS mapping images for Pb, S, and O are shown, respectively. The NP, that is composed of Pb and S, sits on top of the interfacial layer which contains a high concentration of oxygen formed mainly by an incomplete etching. We have previously shown that the thickness of the interfacial layer is usually inhomogeneous under the NPs [24]. Part of the NPs can be found directly in contact with the InP surface, enabling an epitaxial relationship between the InP and PbS lattices. As the crystallographic order of the NPs is influenced by the crystal structure of the substrate, it explains the uniformity of the NP orientation on the etched InP surface. While the oxide layer can be suppressed by a passivation of the InP surface prior to the growth of the NPs [24], it is the presence of the oxide that makes this structure interesting for neuromorphic computing.

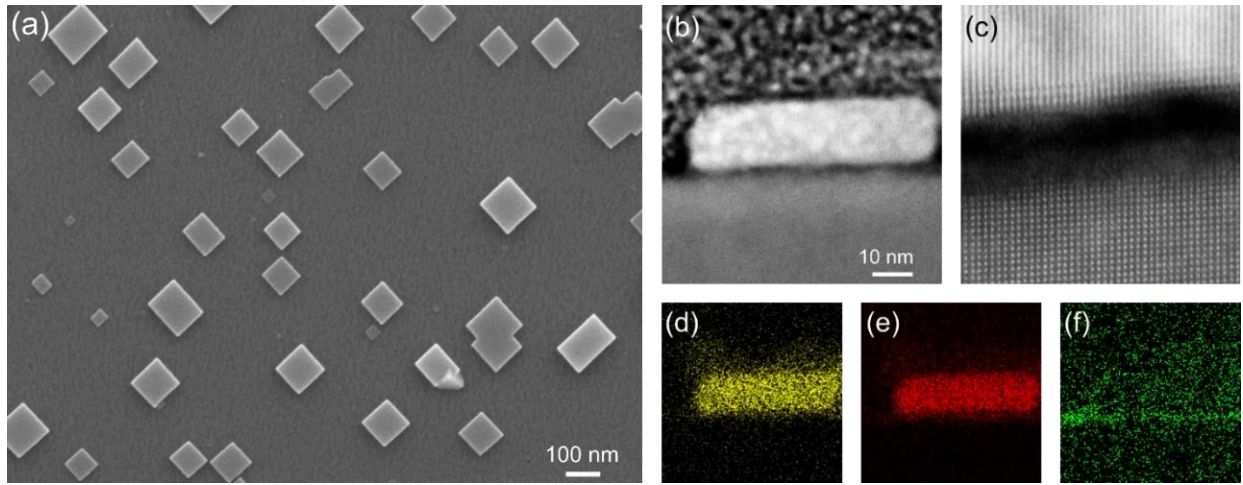

Figure 1 (a) Scanning electron micrograph of PbS nanoplatelets grown on an InP (001) (b) (100) cross-sectional high-angle annular dark-field (HAADF) scanning transmission electron microscopy (STEM) of a PbS nanoplatelet. (c) High resolution (110) cross-sectional STEM image of the PbS/InP interface. (d), (e), (f) Energy dispersive spectroscopy of the heterointerface for Pb, S and O respectively.

Scanning tunneling spectroscopy (STS) was used in order to characterize the transport characteristics of the samples, employing tungsten tips in an ultra-high vacuum ($5 \times 10^{-11}$ Torr) chamber, at 77 K. Sequences of I-V curves are typically acquired on the NPs by applying a voltage ramp from negative towards positive voltages with a duration of 4.6 s and then a faster voltage ramp from positive towards negative voltages (1.7 s). As shown in Figure 2(a), a plateau is observed at positive bias and the onset of the plateau shifts to a higher bias as more and more spectra are acquired, consistent with previous measurements [24]. The initial threshold voltage of the plateau occurs for a Fermi level of the tip that is positioned below the conduction band edge of the PbS NP and indicates the presence of traps in the oxide layer, since without this oxide layer the I-V characteristics were found to be markedly different. If the traps cannot be discharged during the positive to negative voltage sweep, then the potential of the system will be higher and causes the threshold voltage of the plateau to be shifted during the next voltage ramp [25], as seen in Figure 2(a). While previous analyses of the heterostructures indicates a conduction band edge of InP positioned below the conduction band edge of PbS [24] [26], the presence of the oxide layer modifies the band alignment. Based on the literature of trap states found in the native oxide at the surface of InP substrate [27] [28], we suspect the trap states to be positioned in the band gap of InP, rather close to the bottom of the InP conduction band. Then, the transport through the trap states involve a non-radiative recombination process between tunneling electrons and holes in InP. Due to the formation of a depletion layer or an inversion layer in the InP susbtrate (Figure 2(b)), such a process is usually not efficient, suggesting a long detrapping time that accounts for the shift of the plateau towards higher bias with increasing number of spectra.

In order to confirm this hypothesis, we set the measurements conditions so that the traps states were partially filled, so that the plateau is shifted to a high bias and its edge is rather smooth. Then, I-V measurements were acquired with increasing set-point currents corresponding to a reduction of the tunneling barrier between the STM tip and the NP. As shown in Figure 2(c) for another NP, the voltage threshold of the plateau decreases with increasing set-point currents. This shift takes along with the formation of a steeper onset, reaching a saturation regime above 600 pA. This behavior is the signature of a two-step process: the first one is the tunneling of the electron through the PbS NP into the trap state, the second is the process one to leave the traps. This second process can limit the current, leading to the observed plateaus, in accordance with similar results obtained for a single quantum level [29].

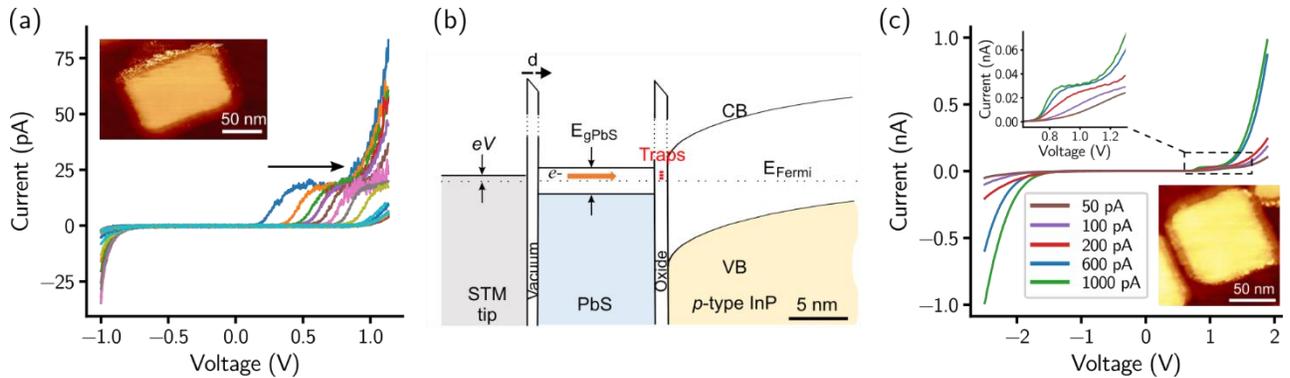

Figure 2 (a) Sequences of twenty tunneling spectra acquired on the PbS nanoplatelet shown in the inset. The arrow points the voltage shift of the onset of the plateau that is induced by the repetition of the spectra. (b) Band diagram of the double tunneling junction between the tip, the PbS nanoplatelet and the inversion layer of the p-type InP substrate. Trap states are positioned in the oxide potential barrier. The horizontal orange arrows indicate the direct tunneling of electrons from the tip into these states. The dashed arrow indicates the variable thickness of the vacuum barrier depending on the set-point current. The InP valence and conduction bands and the applied bias are labeled VB, CB and V respectively. (c) Scanning tunneling spectroscopy of a PbS nanoplatelet measured with increasing set-point currents corresponding to decreasing tip-sample distances. Upper inset: Zoom into the voltage range where a plateau is seen. Lower inset: STM image of the PbS nanoplatelet.

If now the duration of the spectra acquisition is changed, with a voltage ramp from negative to positive bias applied during periods varying from 0.07 s to 4.6 s and followed by a faster voltage ramp from positive towards negative voltages, an increase of the measured currents is also observed, as demonstrated by the graphs in Figure 3(a). In addition, the plateau voltage shift is minimized by a waiting time of 60 s between each spectrum acquisition. The inset in Figure 3(a) displays the relationship between the measured current and the acquisition time. It yields an inversely proportional behavior. A similar trend is also obtained by performing time-resolved measurements. In this case, the initial bias used for the feedback conditions is negative (-0.4 V) allowing the trap to be discharged. Then a pulse $V_1 = +0.3$ V is applied and the current $I_1$ is acquired for 30 ms after a delay $\tau$, before the voltage is reset to the initial negative value to completely discharge the trap. This sequence is repeated by increasing $\tau$ stepwisely. The scheme of the measurement is illustrated in Figure 3(b). We can observe from the results in Figure 3(c) that the measured current $I_1$ is inversely proportional to the duration of the pulse. This quantity is related to the sampling rate of standard measurements, which in turn affects the acquisition time from the measurements of Figure 3(b). With this, it is clear that this structure, meso-scopic in size, manifests a learning and memory functionality where a threshold voltage increases after each measurement, equivalent to the corresponding behaviors of a neuronal unit. Moreover, there is a memory fading equivalence as reflected in the observation of a short relaxation time or a decreased resistivity if measurements are performed with a high enough frequency.

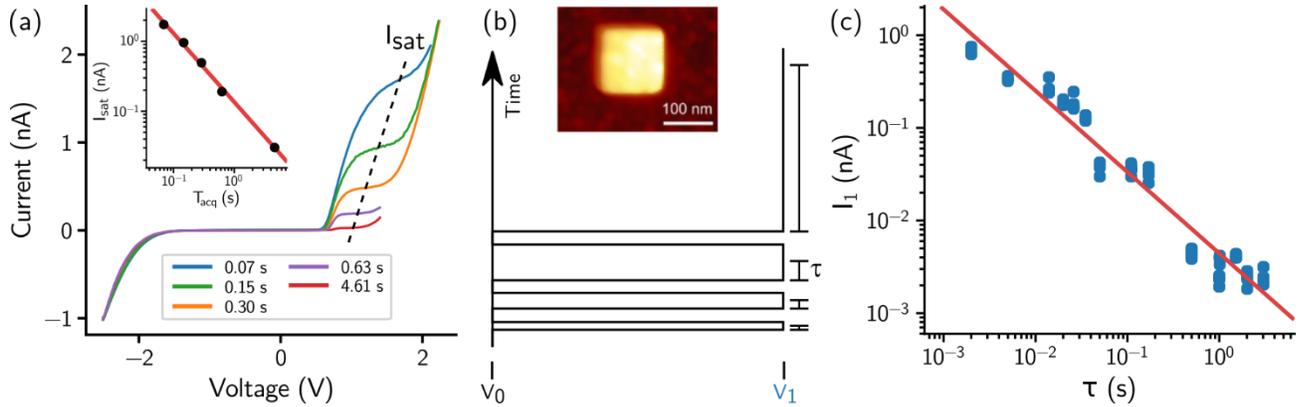

Figure 3 (a) Spectra measured with different acquisition times. The influence of the acquisition time on the magnitude of the current measured in the middle of the plateau in shown in the inset. (b) Time-resolved I (V) spectroscopy measurement scheme, that was used to probe the NP shown in the inset. The sample is negatively bias at $V_0$. Then a pulse is applied to set the sample at a positive voltage $V_1$. $\tau$ is the delay used before the measurement of the current during the pulse. (c) Influence of the delay time on the magnitude of the current $I_1$ measured at the voltage $V_1$ of +0.3 V for a reset bias $V_0$ of -0.4 V.

## 3. SET modelling

As an attempt to understand the transport behavior observed from the PbS NPs on a p-type InP substrate, we consider the structure shown in Figure 2(b) and a two-step process for electron transport through the oxide barrier: the first is the tunneling of the electron from the STM tip through the PbS NP and into a trap state, and the second is the process to leave the trap.

We find one can use a Single Electron Tunneling (SET) abstract model to describe this physical system as an arrangement of resistors and capacitors, where each tunneling junction is represented as a resistor serially connected to a second resistor in parallel to a capacitor. As the structure is charged with one electron, we can write:

$$E_e = \frac{e^2}{2C} \quad (1)$$

In this equation, $e$ stands for the electron charge, $C$ the total capacitance, and $E_e$ the energy increase corresponding to one added electron.

This energy increase is equivalent to a quantum state filled by an electron that is able to tunnel through the first barrier. Since the state is changed, subsequent electrons must possess a higher energy ($> E_e$) to tunnel through the first potential barrier, preventing the flow of a higher current if a high enough energy is not supplied. This effect is known as Quantum Blockade [6].

If this charging effect is indeed compatible with the observed results, there are two conditions that must be met. The first one is in regard to the discharging time $\tau_{RC}$:

$$\tau_{RC} = RC \quad (2)$$

To be measurable, $\tau_{RC}$ must be much greater than the quantum state lifetime, that obeys the uncertainty principle:

$$\Delta t \Delta E \geq \frac{\hbar}{2} \rightarrow \Delta t \geq \frac{\hbar C}{e^2} = \quad (3)$$

$$\tau_{RC} \gg \Delta t \rightarrow R \gg \frac{\hbar}{e^2} \approx 4k\Omega \quad (4)$$

And so, we now have a limit for the RC circuit time constant, or, more specifically, the circuit resistance, which should be much higher than 4 kΩ. The second condition is that the energy required for one electron must be higher than thermal noise at a certain temperature T:

$$E = \frac{e^2}{2C} > k_B T \rightarrow T < \frac{e^2}{2k_B C}, \quad (5)$$

where $k_B$ stands for the Boltzmann's constant. This expression limits the temperature at which these effects can be observed, depending on the RC circuit capacitance.

To verify that the observed behavior is consistent with a double-junction SET [30] [31], a least-squares fitting was performed on I-V curves slowly obtained (steady-state) from completely discharged samples, considering a combination of diode curves and sigmoid functions as a fitting model. From the plateau onset, width, slope, and edge curvature, it is possible to obtain the values for the capacitance and resistance of each junction [32]. For the tip-vacuum-PbS junction, the values obtained were $C_1 = 5 \times 10^{-19} F$ and $R_1 = 20\ G\Omega$. For the PbS-Oxide-InP junction, the values obtained were $C_2 = 2 \times 10^{-19} F$ and $R_2 = 4\ G\Omega$. Although smaller than one could expect, the value for $C_2$ is not surprising, considering that the capacitance value does not scale with the area in the nanoscale [33], but with the average dimension (diagonal), due to the possible formation of dead layers [34]. Another consideration is that there is also depletion around the perimeter of the nanoplatelet, which extends to the InP layer, as the surface states cause only the center zone to be active to external bias. Example I-V curves and the respective SET I-V model results are shown in Figure 4(a), where the SET circuit schematic is shown in the inset. The resistance values are clearly higher than the condition from (4) requires. The limiting temperature from (5) for both cases is greater than 300K, and thus perhaps also indicative of the system's potential in neuromorphic applications.

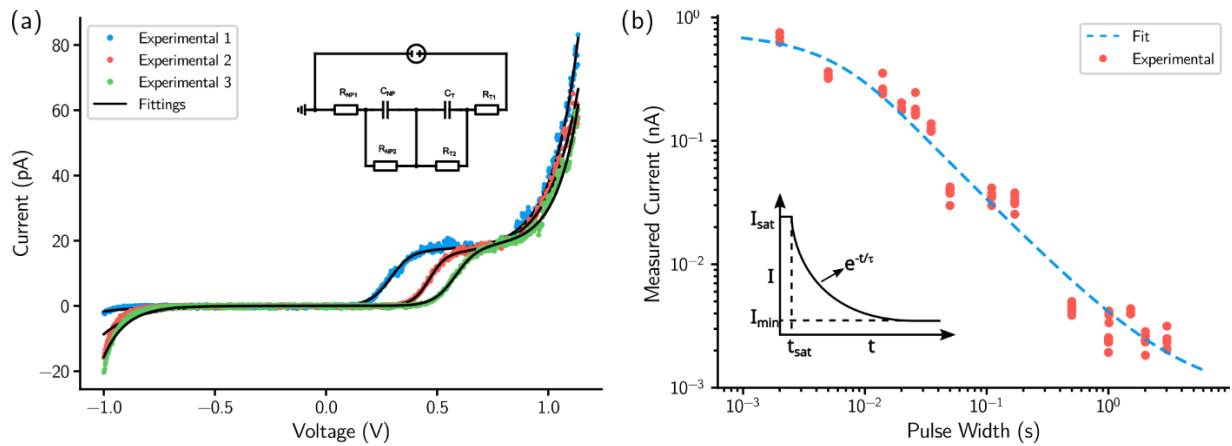

Figure 4 (a) Examples of fitted I – V curves using a SET model composed of capacitors and resistors (inset). (b) Time-dependent current response to voltage pulses modelled as the integration of an exponentially decaying response (inset).

While being in good agreement with the SET phenomenon, the system has some particularities that need to be addressed. First, the asymmetry is evident, which can be easily explained by the highly asymmetric structure. When electrons move in the InP-Oxide-PbS-Vacuum-STM direction (negative bias), the trap states are certainly not accessible to the valence band electrons in InP, as seen in Figure 2(b). Second, we only see one "stair" in the Coulomb staircase in the positive bias regime. This single step is explained by the particular oxide trap tunneling mechanism, the small effective conductive area of the junction associated with the PbS nanoplatelet, which causes the staircase effect to be more pronounced. The likely presence of multiple traps is equivalent to that of many smaller "islands". When a high enough voltage is applied, instead of revealing another step, the resultant current is already higher than the presumed next plateau current, since the tunneling now happens with a higher energy than the trap levels.

The impulse response shown in Figs. 3(a) and 3(c) can be modeled as a fast charging high-current response followed by an exponential decay. As such, we can treat the data obtained as the integration of this response during the voltage pulse, as shown in Figure 4(b), obtaining 700 pA for the highest current response, 1 pA for the minimum current, a duration of 200 ns for the constant high-current response, and an exponential decay time constant (relative to the discharging rate) of 4 ms. As the electrons fill the traps, the number of available paths for subsequent electrons is decreased, causing the decrease in current for longer pulses.

## 4. Applications to neuromorphic computing

While being much smaller in physical size, the PbS/InP structures present many similarities to biological neurons, grown in liquid phase, scalable self-assembling, and amendable to vertical stacking into many layers of interconnected elements; from the current-voltage measurements, it is clear that these devices present memory and memory fading effects, which are fundamental to learning capabilities. Additionally, a non-linear response and thresholding behavior, which can be observed in the experimental results, are important characteristics of an artificial neuron.

With the parameters obtained from the measurements, we can create a model to simulate the response of a PbS/InP device employed in different neuromorphic computing applications, under a variety of conditions. To do this, we use the SET I-V model as a transfer function for the electrical current. From the experimental data, a time-dependent modification of the transfer function can be implemented, resulting in conductivity changes and plateau onset shifts. By mimicking voltage ramps with well determined durations or successive voltage pulses, the behaviors observed during the measurements are successfully replicated, as shown in Figure 5(a) and (b).

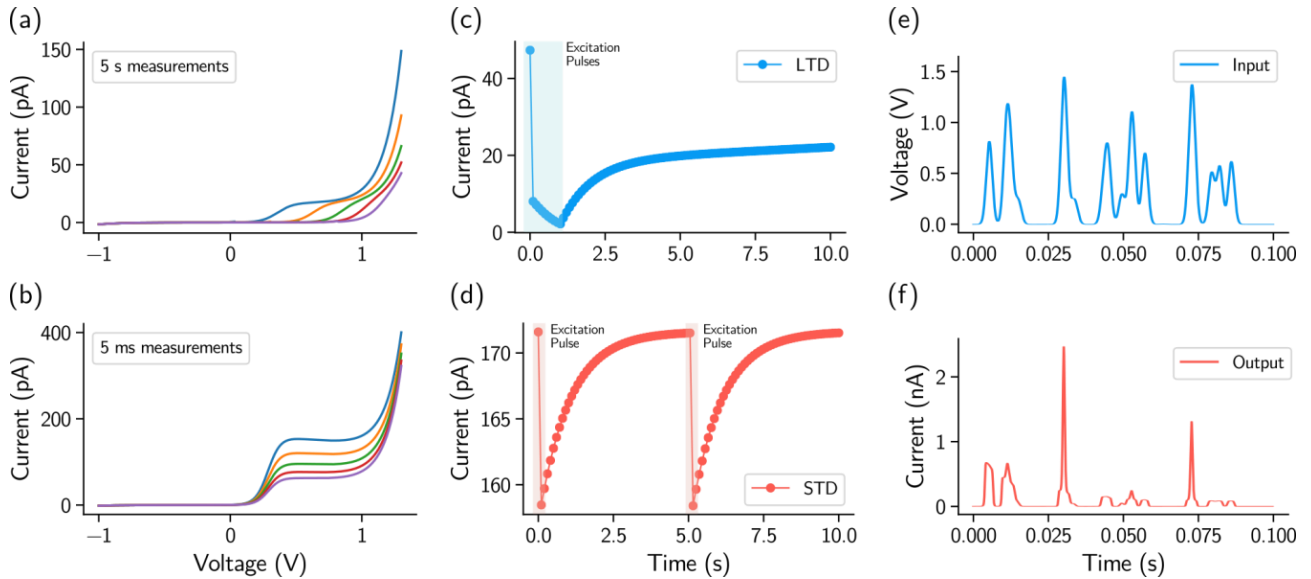

Figure 5 (a) Reproduction of the plateau onset voltage shift after subsequent (long) measurements, obtained using the SET model with fitted parameters. (b) Reproduction of decreasing plateau voltages for longer measurements. (c) Demonstration of long-term depression after a series of 100 ms excitation pulses occurring within the blue area, obtained from the developed model. (d)

Demonstration of short-term depression after one 1 ms pulse, occurring within the red areas. (e) Example of a series of random weighted spikes, analogous to the information a neuron can receive. (f) Output of the modelled neuron to the input in (e).

As a first example, we use this model to demonstrate the capability for synaptic behavior, exhibiting short and long-term plasticity effects. Voltage pulses (~ms), analogous to brain impulses, can be applied to change the device conductivity state, while shorter (~ns) pulses can be used to probe the current response of the device.

Figures 5(c) and (d) shows the emulated current resulting from probe pulses as function of time, for different excitation pulses. When voltage pulse trains (100 ms) are applied, a typical long-term depression effect with time constant in the order of 100 s can be observed in Figure 5(c). The short term effect can be recognized in Figure 5(d) when a single pulse (1 ms) is applied, with a time constant in the order of 1 s.

An additional example of neuromorphic computing applications is shown in Figure 5(e) and (f). Random pulses (analogous to brain spikes originated from neighboring synapses) are added together and used as input to an emulated PbS device. Due to its non-linear response and thresholding characteristics, this artificial neuron provides a response that changes in time (conductivity change and shift in threshold depending on previous impulses), firing new spikes when the threshold is achieved, corresponding to processed information.

Finally, we can also estimate the energy consumption and speed limit of each component based on the experimental results. From the fitted parameters of Figure 4, the voltage pulse width necessary to cause a change in the conductivity is approximately 200 ns, equivalent to a maximum frequency of 5 MHz. Pulses shorter than this will not have any effect on short or long-term conduction characteristics, limiting the processing speed of the devices. Considering this time scale, and an operating point within the highest plateau onset of Figure 3(a) (1 V, 1 nA region), the energy dissipated purely by Joule effect is of the order of 0.2 fJ. This is an estimation of the minimum energy required for an operation on an average PbS/InP device used for neuromorphic computing, a value that is substantially lower than human brain synapses (~pJ) [23], demonstrating the potential of this material for such applications.

## 5. Conclusion

In this work, we have demonstrated how liquid-phase grown PbS nanoplatelets on InP can display characteristics that are advantageous for neuromorphic computing, originating from SET mechanisms. Presenting thresholding behavior and both short- and long-term memory effects that affect its inherently non-linear current response, devices based on this PbS/InP structure can be employed in neuromorphic computing applications, acting as both synapses (weighted connections) and spiking elements. A large number of elements with different (random) characteristics can be connected to provide a larger scale neural-network made entirely of these structures, which can be easily fabricated on large-scale with a liquid-phase growth technique. In comparison with the human brain, this material has the potential to present faster and more energy-efficient operations, in the range of 200 ns and 0.2 fJ per synaptic activity.